\documentclass[preprintnumbers,amsmath,amssymb,nofootinbib,superscriptaddress,floatfix,onecolumn,reprint,
groupedaddress,
eqsecnum,%eqsecnum自动按照章节编号，章节符号是阿拉伯数字。
amsfonts,aps]{revtex4}

\pdfoutput=1
\usepackage[normalem]{ulem}%下划线宏包：\uline{my}可以在正文中my下面加下划线；\uuline{my} 可以在正文中my下面加双下划线；\uwave{my}下面加波浪线；\sout{my}将这个词用水平线删除；\xout{my}将这个词用斜线删除；
\usepackage{graphicx}
\usepackage[colorlinks,citecolor=blue,urlcolor=blue,linkcolor=blue]{hyperref}% 超链接宏包

\usepackage{subfigure}%关于图形的宏包，其说明文件已经和该论文放在一起
\usepackage{latexsym}
\usepackage{amssymb}
\usepackage{mathrsfs}
\usepackage{amsmath}
\usepackage{amsthm}
\usepackage{color}
\usepackage{dcolumn}% Align table columns on decimal point
\usepackage{bm}% bold math
\definecolor{dyellow}{rgb}{1.,0.8,.0}
\definecolor{myblue}{rgb}{.1,.1,.7}
\definecolor{dcyan}{rgb}{.0,.6,.6}
\definecolor{dmagenta}{rgb}{0.6,0.0,0.6}
\definecolor{brown}{rgb}{0.6,0.2,0.}
\definecolor{darkblue}{rgb}{.0,.0,0.5}
\definecolor{darkred}{rgb}{0.75,0.0,0.0}
\definecolor{orange}{rgb}{1.,.6,.0}
\definecolor{dorange}{rgb}{0.8,.4,.0}
\definecolor{darkgreen}{rgb}{0.0,0.6,0.0}
\definecolor{purple}{rgb}{.4,.0,.4}
\definecolor{grey}{rgb}{0.5,0.5,0.5}
\def\black{\color{black}}

\begin{document}
\hyphenpenalty=1000%等号后数值越大，可以更可能减少换行时断字太多的问题
%\preprint{APS/123-QED}
\title{Multipole analysis on stationary massive vector and symmetric tensor fields with irreducible Cartesian tensors}

\author{Bofeng Wu}%\affiliation{\NEU}
\affiliation{Department of Physics, College of Sciences, Northeastern University, Shenyang 110819, China}
%Department of Physics, College of Sciences, Northeastern University, Shenyang 110819, China}

%\author{Chao-Guang Huang}%\footnote{Corresponding author}}
%\email{huangcg@ihep.ac.cn}
%\affiliation{Institute of High Energy Physics, Chinese Academy of Sciences, Beijing, 100049, China}

%\date{\today}
\begin{abstract}
\centerline{\textbf{ABSTRACT}}
\bigskip
The multipole expansions for massive vector and symmetric tensor fields
in the region outside spatially compact stationary sources are obtained by using the symmetric and trace-free formalism in terms of
the irreducible Cartesian tensors, and the closed-form expressions for the source multipole moments are provided.
The expansions show a Yukawa-like dependence on the massive parameters of the fields, and the integrals of the stationary source multipole moments are all modulated by a common radial factor. % related to  the source distribution.
For stationary massive vector field, there are two types of ``magnetic'' multipole moments, among which one is the generalization of that of the magnetostatic field, and another, being an additional set of multipole moments of the stationary massive vector field, can not be transformed away. As to the stationary massive symmetric tensor field, its multipole expansion is presented when the trace of its spatial part is specified, where besides the counterparts of the mass and spin multipole moments of massless symmetric tensor field, three additional sets of multipole moments also appear. The multipole expansions of the tensor field under two typical cases are discussed, where it is shown that if the spatial part of the tensor field is trace-free, the monopole and dipole moments in the corresponding expansion will vanish.
\end{abstract}
%\pacs{04.50.Kd,  04.25.-g, 04.25.Nx}
\maketitle
\section{Introduction}\label{sec1}
A massive vector field is usually used to describe a massive particle with spin-1~\cite{Kurt,Parker}, and mathematically, the field equations of a massive vector field $A^{\mu}$ can be written as
\begin{subequations}
\begin{eqnarray}
\label{equ1.1a}&&\square A^{\mu}-m_{A}^{2}A^{\mu}=-4\pi j^{\mu},\\
\label{equ1.1b}&&\partial_{\mu}A^{\mu}=0,
\end{eqnarray}
\end{subequations}
where $\square:=\eta^{\mu\nu}\partial_{\mu}\partial_{\nu}$ with $\eta^{\mu\nu}$ as the Minkowskian metric in a Minkowski spacetime and $\partial_{\mu}$ as the partial derivative with respect to the Minkowskian coordinate, $m_{A}$ is the massive parameter of the field, and $j^{\mu}$ is the external source. Although Eq.~(\ref{equ1.1a}) formally coincides with the Lorenz gauge condition in Maxwell's electromagnetic theory, it is important to realize that due to the existence of $m_{A}$,
Eq.~(\ref{equ1.1b}) arises dynamically, i.e. as consequence of the the continuity
equation satisfied by $j^{\mu}$, and there is no longer the local gauge symmetry for $A^{\mu}$.
As a contrast, a massive spin-2 particle could be depicted by a massive symmetric tensor field (MSTF) $h^{\mu\nu}$, and
its field equations can be mathematically assumed to be
\begin{subequations}
\begin{eqnarray}
\label{equ1.2a}&&\square h^{\mu\nu}-m_{h}^{2}h^{\mu\nu}=-4\pi T^{\mu\nu},\\
\label{equ1.2b}&&\partial_{\mu}h^{\mu\nu}=0,
\end{eqnarray}
\end{subequations}
where $m_{h}$ is the massive parameter of the field, $T^{\mu\nu}$ is the external symmetric source, and the constraint~(\ref{equ1.2b}) is compatible with the conservation of $T^{\mu\nu}$. The above two equations have a wide range of applications in the models of massive gravity~\cite{Maggiore2008,higuchi1989massive,hinterbichler2012theoretical,gambuti2021fierz}. Unlike the case in linearized General Relativity (GR), the term $m_{h}^{2}h^{\mu\nu}$ in Eq.~(\ref{equ1.2a}) results in that $h^{\mu\nu}$ does not have the usual massless gravity gauge symmetry.
Both of massive vector and symmetric tensor fields have important applications in theoretical physics~\cite{clough2022problem, mikki2021proca, erofeev2020dynamic, gambuti2020note, asano2019minimal, Cardoso:2019mes}, and therefore, it is very necessary to explore the solutions to their field equations.

One of the most significant ways to describe the external field of the source localized in a finite region of space is the multipole expansion. The symmetric and trace-free (STF) formalism in terms of the irreducible Cartesian tensors, developed by Thorne~\cite{Thorne:1980ru}, Blanchet, and Damour, and Iyer~\cite{Blanchet:1985sp,Blanchet:1989ki,Damour:1990gj}, is one useful method with respect to the multipole expansion. In view of the important applications of the above massive vector and symmetric tensor fields in physics, the multipole analysis on them in the region outside spatially compact sources is very necessary. In Ref.~\cite{Damour:1990gj},
the relativistic multipole expansions for massless vector and symmetric tensor fields are obtained using the STF formalism, and the corresponding source multipole moments are derived. By following the same approach, we can also make a multipole analysis on the  massive vector and symmetric tensor fields.

The starting point of this study is the multipole expansion for the massive scalar field, namely the Klein-Gordon field.
When the source is time-independent, the field equation of the Klein-Gordon field reduces to the screened Poisson equation, and since
the Green's function of this equation is easier to handle, the derivation in such case can be greatly simplified.
%In view of this, only stationary sources are considered in the present paper.
In Ref.~\cite{Wu:2017vvm},  the multipole expansion for the Klein-Gordon field with a spatially compact stationary source is derived, and the closed-form expressions of the source multipole moments are provided. In this paper, using the STF formalism, we will make use of this result to make a multipole analysis on the massive vector and symmetric tensor fields in the region outside spatially compact stationary sources.

The derivation can be performed by following the conventional method in Refs.~\cite{Blanchet:1985sp,Damour:1990gj}, and compared with the results for stationary massless vector and symmetric tensor fields, the multipole expansions for stationary massive fields show a Yukawa-like dependence on the massive parameters of the fields, and the integrals of the stationary source multipole moments are all modulated by a common radial factor. % related to the source distribution.
For stationary massive vector field $A^{\mu}$, the multipole expansion of $A^{0}$ field and the ``electric'' multipole moments are compatible with the multipole expansion of the scalar potential of the electrostatic field presented in Ref.~\cite{Damour:1990gj}, and it is shown that outside the source region, the stationary $A^{0}$ field can be equivalently generated by the source built from $\delta$-function, which is referred to as the skeleton of the stationary $A^{0}$ field. As a contrast, the multipole expansion of $A^{i}$ field is different from that of the vector potential of the magnetostatic field because two sets of ``magnetic'' multipole moments appear in the expansion, where one of them is the generalization of that of the magnetostatic field, and another one, as an additional set of multipole moments of the stationary massive vector field, can not be transformed away because there is no the local gauge symmetry for $A^{\mu}$. Moreover, it should be pointed out that for the massive vector field, both of the two types of ``magnetic'' monopole moments always vanish.

As to the stationary massive symmetric tensor field $h^{\mu\nu}$, its multipole expansion is also presented. In the expansion, besides the counterparts of the mass and spin multipole moments of massless symmetric tensor field, there are three additional sets of multipole moments. Similarly, since $h^{\mu\nu}$ has no the usual massless gravity gauge symmetry, these three additional sets of multipole moments can also not be transformed away. The derivation indicates that the multipole expansion for $h^{ij}$ is dependent on its trace, and hence, for future application, the expression of $h^{kk}$ also needs to be given. The results under the cases of $h^{kk}=0$ and $h^{kk}=h^{00}$ are provided in the present paper. It is shown that when $h^{kk}=0$, the monopole and dipole moments in the expansion of $h^{ij}$ vanish, which is compatible with the general form of the STF-tensor spherical harmonics expansion for a trace-free tensor field of ``spin'' 2 on the unit sphere centered at the coordinate origin~\cite{Blanchet:1985sp}.

The multipole expansion for stationary massive symmetric tensor field, as the external solution to Eqs.~(\ref{equ1.2a}) and (\ref{equ1.2b}) for any spatially compact stationary source, describes the effects of the source at all orders, so it must have important applications in the models of massive gravity.
In addition, it also plays an important role in the alternative theories of gravity. In Ref.~\cite{Wu:2022mna}, the metric for the external gravitational field of a spatially compact stationary source is discussed in $F(X,Y,Z)$ gravity, a generic fourth-order theory of gravity,  where $X:=R$ is the Ricci scalar, $Y:=R_{\mu\nu}R^{\mu\nu}$ is the quadratic contraction of two Ricci tensors,  $Z:=R_{\mu\nu\rho\sigma}R^{\mu\nu\rho\sigma}$ is the quadratic contraction of two Riemann tensors, and $F$ is a general function of $X, Y,$ and $Z$. In the derivation, the linearized gravitational field equations of $F(X,Y,Z)$ gravity are transformed into d'Alembert equation and Klein-Gordon equations with external sources by imposing a new type of gauge condition. Specifically, it is shown that there exists a symmetric tensor $P^{\mu\nu}$, constructed from the linearized Ricci tensor $R^{\mu\nu(1)}$ and the linearized Ricci scalar $X^{(1)}$,  satisfying
\begin{subequations}
\begin{eqnarray}
\label{equ1.3a}&&\square P^{\mu\nu}-m_{p}^{2}P^{\mu\nu}=-4\pi \iota S^{\mu\nu},\\
\label{equ1.3b}&&\partial_{\mu}P^{\mu\nu}=0
%\label{equ1.3c}&&P^{\mu}_{\phantom{\mu}\mu}=0
\end{eqnarray}
\end{subequations}
outside the source, where the massive parameter $m_{p}$ is defined by the coefficients of $X, Y,$ and $Z$ when $F(X,Y,Z)$ is expressed as a power series, $\iota$ is a constant, and the symmetric tensor $S^{\mu\nu}$ is relevant to the energy-momentum tensor of the source living in a Minkowski spacetime. The tensor $P^{\mu\nu}$ presents a massive propagation in $F(X,Y,Z)$ gravity~\cite{Stabile:2010mz,stabile2015}, and one can obtain the metric for the gravitational field outside a spatially compact stationary source in $F(X,Y,Z)$ gravity only after the solution to Eqs.~(\ref{equ1.3a}) and (\ref{equ1.3b})  is given.

In Ref.~\cite{Wu:2022mna}, since the multipole expansions of the time-space and spatial components of $P^{\mu\nu}$ are not derived, the metric provided in Ref.~\cite{Wu:2022mna} is not expanded in terms of the irreducible Cartesian tensors, which greatly limits the application of the result. Comparing Eqs.~(\ref{equ1.3a}) and (\ref{equ1.3b}) with Eqs.~(\ref{equ1.2a}) and (\ref{equ1.2b}), respectively, we find that by applying the result derived in the present paper, the complete multipole expansion for the stationary $P^{\mu\nu}$ will be easily gained, and as a consequence, the metric, presented in the form of the multipole expansion,  for the external gravitational field of a spatially compact stationary source will also be obtained in $F(X,Y,Z)$ gravity. According to the result in this paper,  besides the counterparts of the mass and spin multipole moments in the linearized GR, there should be three additional sets of multipole moments appearing in the expansion of the metric, and when the metric is applied to some specific phenomenon in practice, the effects of those terms associated with the additional moments can be analyzed. For instance, for a gyroscope moving around the source in geodesic motion, one is able to use the metric to derive its spin's angular velocity of precession, and by following the conventional method in Ref.~\cite{MTW1973}, the precessional angular velocity in GR would be corrected by those terms associated with the additional moments. Then, by comparing these results with the data of the gyroscopic experiment, e.g., Gravity Probe B (GP-B), the effects of the additional moments in $F(X,Y,Z)$ gravity will be obtained. Similarly, the metric can also be applied to the anomalous perihelion advance of Mercury, the gravitational redshift of light, and the light bending, etc. By comparing the theoretical results with the experimental or observational data, the further effects of the additional moments in $F(X,Y,Z)$ gravity will also be derived. It could be expected that in the future, more and more applications of the results in the present paper will be found.
\black

This paper is organized as follows. In Sec.~\ref{sec2}, the STF formalism and the multipole expansion for a stationary Klein-Gordon field are briefly reviewed. In Sec.~\ref{sec3}, a multipole analysis on stationary massive vector field is made by using the STF formalism. In Sec.~\ref{sec4}, the STF formalism is extended to deal with the stationary massive symmetric tensor field. In Sec.~\ref{sec5}, the conclusions and the related discussions are presented.
%The international system of units is used
Throughout this paper, when the notation is concerned, the Greek letters denote spacetime indices and range from 0 to 3, whereas the Latin letters denote space indices and range from 1 to 3. The repeated indices within a term represent that the sum should be taken over.
%%%%%%%%%%%%%%%%%%%%%%%%%%%%%%%%%%%%%%%%%%%%%%%%%%%%%%%
%%%%%%%%%%%%%%%%%%%%%%%%%%%%%%%%%%%%%%%%%%%%%%%%%%%%%%%
%%% The first section                              %%%%
%%%%%%%%%%%%%%%%%%%%%%%%%%%%%%%%%%%%%%%%%%%%%%%%%%%%%%%
%%%%%%%%%%%%%%%%%%%%%%%%%%%%%%%%%%%%%%%%%%%%%%%%%%%%%%%
\section{Preliminary}\label{sec2}
\subsection{Relevant notations and formulas in the STF formalism~\label{Sec:STF}}
The notation in this paper is the same as that in Refs.~\cite{Wu:2017vvm,Wu:2018hjx,Wu:2018jve,Wu:2021uws}. In a Minkowski spacetime with signature $(-,+,+,+)$, the STF part of a Cartesian tensor $A_{I_{l}}:=A_{i_{1}i_{2}\cdots i_{l}}$ is denoted by
\begin{eqnarray}
\label{equ2.1}
\hat{A}_{I_{l}}:=A_{\langle I_{l}\rangle}=A_{\langle i_{1}i_{2}\cdots i_{l}\rangle}:=\sum_{k=0}^{\left[\frac{l}{2}\right]}c_{k}\delta_{(i_{1}i_{2}}\cdots\delta_{i_{2k-1}i_{2k}}S_{i_{2k+1}\cdots i_{l})a_{1}a_{1}\cdots a_{k}a_{k}},
\end{eqnarray}
where $\left[l/2\right]$ representing the integer part of $l/2$, $\delta_{ij}$ denoting the Kronecker symbol,
\begin{equation}
\label{equ2.2}c_{k}:=(-1)^{k}\frac{(2l-2k-1)!!}{(2l-1)!!}\frac{l!}{(2k)!!(l-2k)!},
\end{equation}
and
\begin{equation}\label{equ2.3}
S_{I_{l}}:=A_{(I_{l})}=A_{(i_{1}i_{2}\cdots i_{l})}:=\frac{1}{l!}\sum_{\sigma} A_{i_{\sigma(1)}i_{\sigma(2)}\cdots i_{\sigma(l)}}
\end{equation}
is its symmetric part with $\sigma$ running over all permutations of $(12\cdots l)$. Denote $(x^{\mu})=(ct,x_{i})$ and $(ct,r,\theta,\varphi)$ as Minkowskian coordinates and the corresponding spherical coordinates, and then, there are
\begin{equation}\label{equ2.4}
x_{1}=r\sin{\theta}\cos{\varphi},\ x_{2}=r\sin{\theta}\sin{\varphi},\ x_{3}=r\cos{\theta}.
\end{equation}
%with $r=\sqrt{x_{i}x_{i}}$.
Let $\partial_{i}:=\partial/\partial x_{i}$ be the coordinate basis vectors, and the radial vector and the unit radial vector are $\boldsymbol{x}=x_{i}\partial_{i}$ and $\boldsymbol{n}=n_{i}\partial_{i}=(x_{i}/r)\partial_{i}$, respectively.
With $x_{i}$ and $n_{i}$, the tensor products of $l$ radial and unit radial vectors are
\begin{eqnarray}
\label{equ2.5}X_{I_{l}}&=&X_{i_{1}i_{2}\cdots i_{l}}:= x_{i_{1}}x_{i_{2}}\cdots x_{i_{l}},\\
\label{equ2.6}N_{I_{l}}&=&N_{i_{1}i_{2}\cdots i_{l}}:= n_{i_{1}}n_{i_{2}}\cdots n_{i_{l}},
\end{eqnarray}
and they satisfy
\begin{equation}\label{equ2.7}
X_{I_{l}}=r^l N_{I_{l}}.
\end{equation}

In the STF formalism, one important result is that any Cartesian tensor $T_{I_{p}}$ can be decomposed into a finite sum of terms of the type $\gamma_{I_{p}}^{J_{l}}\hat{R}_{J_{l}}$, where $\gamma_{I_{p}}^{J_{l}}$ is a tensor invariant under the group of proper rotations $\text{SO(3)}$, and $\hat{R}_{J_{l}}$ is an irreducible STF $l$ tensor $(l\leq p)$~\cite{Blanchet:1985sp,Damour:1990gj,coope1970irreducible,coope1965irreducible}. This assertion can directly be proven by induction if one uses the following formula:
\begin{equation}\label{equ2.8}
A_{i}\hat{T}_{I_{l}}=\hat{R}^{(+)}_{iI_{l}}+\frac{l}{l+1}\epsilon_{ai\langle i_{l}}\hat{R}^{(0)}_{i_{1}\cdots i_{l-1}\rangle a}+\frac{2l-1}{2l+1}\delta_{i\langle i_{l}}\hat{R}^{(-)}_{i_{1}\cdots i_{l-1}\rangle},
\end{equation}
where
\begin{subequations}
\begin{eqnarray}
\label{equ2.9a}&&\hat{R}^{(+)}_{I_{l+1}}:=A_{\langle i_{l+1}}\hat{T}_{i_{1}\cdots i_{l}\rangle},\\
\label{equ2.9b}&&\hat{R}^{(0)}_{I_{l}}:=A_{a}\hat{T}_{b\langle i_{1}\cdots i_{l-1}}\epsilon_{i_{l}\rangle ab},\\
\label{equ2.9c}&&\hat{R}^{(-)}_{I_{l-1}}:=A_{a}\hat{T}_{ai_{1}\cdots i_{l-1}}
\end{eqnarray}
\end{subequations}
with $\epsilon_{ijk}$ as the Levi-Civita symbol. One particular case of Eq.~(\ref{equ2.8}) is
\begin{equation}\label{equ2.10}
n_{i}\hat{N}_{I_{l}}=\hat{N}_{iI_{l}}+\frac{l}{2l+1}\delta_{i\langle i_{l}}\hat{N}_{i_{1}\cdots i_{l-1}\rangle}.
\end{equation}
Another important result in the STF formalism is that any scalar function $f(\theta,\phi)$ on the unit sphere centered at the coordinate origin can be expanded in powers of the unit radial vector $\boldsymbol{n}$~\cite{Thorne:1980ru,Blanchet:1985sp}, namely,
\begin{equation}\label{equ2.11}
f(\theta,\phi)=\sum_{l=0}^{\infty}\hat{\mathcal{F}}_{I_{l}}\hat{N}_{I_{l}}.
\end{equation}
Here, the STF tensor coefficients $\hat{\mathcal{F}}_{I_{l}}$ are unique, and one is able to obtain
\begin{equation}\label{equ2.12}
 \hat{\mathcal{F}}_{I_{l}}=\frac{(2l+1)!!}{4\pi l!}\int d\Omega \hat{N}_{I_{l}}f(\theta,\phi)
\end{equation}
by virtue of the equality
\begin{eqnarray}
\label{equ2.13}\int\Big(\hat{A}_{I_{l}}\hat{N}_{I_{l}}\Big)\Big(\hat{B}_{J_{l'}}\hat{N}_{J_{l'}}\Big)d\Omega=\frac{ 4\pi l!}{(2l+1)!!}\hat{A}_{I_{l}}\hat{B}_{I_{l}}\delta_{l\,l'}
\end{eqnarray}
where $\hat{A}_{I_{l}}$ and $\hat{B}_{J_{l'}}$ are any two STF tensors, and $d\Omega$ is the element of solid angle.

Let $\partial_{I_{l}}=\partial_{i_{1}i_{2}\cdots i_{l}}:=\partial_{i_{1}}\partial_{i_{2}}\cdots\partial_{i_{l}}$, there is
\begin{eqnarray}
\label{equ2.14}&&\hat{\partial}_{I_{l}}=\sum_{k=0}^{\left[\frac{l}{2}\right]}c_{k}\delta_{(i_{1}i_{2}}\cdots\delta_{i_{2k-1}i_{2k}}
\partial_{i_{2k+1}\cdots i_{l})}\left(\nabla^2\right)^k,
\end{eqnarray}
and related formulas of direct use in later sections are
\begin{eqnarray}
\label{equ2.15}&&\hat{\partial}_{I_{l}}\left(\frac{F(r)}{r}\right)=\hat{N}_{I_{l}}\sum_{k=0}^{l}\frac{(l+k)!}{(-2)^{k}k!(l-k)!}
\frac{\partial_{r}^{l-k}F(r)}{r^{k+1}},\\
\label{equ2.16}&&\partial_{i\langle I_{l}\rangle}:=\partial_{i}\hat{\partial}_{I_{l}}=\hat{\partial}_{iI_{l}}+\frac{l}{2l+1}\delta_{i\langle i_{l}}\hat{\partial}_{i_{1}\cdots i_{l}\rangle}\nabla^{2},
\end{eqnarray}
where $\nabla^2=\partial_{a}\partial_{a}$ is the Laplace operator, $\partial_r^{l-k}$ is the $(l-k)$-th derivative with respect to $r$, and Eq.~(\ref{equ2.16}) can be derived by following the proof of Eq.~(\ref{equ2.10}).
%%%%%%%%%%%%%%%%%%%%%%%%%%%%%%%%%%%%%%%%%%%%%%%%%%%%%%%
%%%%%%%%%%%%%%%%%%%%%%%%%%%%%%%%%%%%%%%%%%%%%%%%%%%%%%%
\subsection{Multipole expansion for a stationary Klein-Gordon field~\cite{Wu:2017vvm}~\label{Sec:K-Gfield}}
In this subsection, we will briefly review the multipole expansion for a stationary Klein-Gordon field. Mathematically, the field equation of a Klein-Gordon field $V(t,\boldsymbol{x})$ with an external source $S(t,\boldsymbol{x})$ is~\cite{greiner}
\begin{eqnarray}
\label{equ2.17}\square V-m^2V=-4\pi S
\end{eqnarray}
with $m$ as the massive parameter of the field. For a stationary source $S(\boldsymbol{x})$ that is spatially compact, the field equation of $V(\boldsymbol{x})$ reduces to the screened
Poisson equation:
\begin{eqnarray}
\label{equ2.18}\nabla^2V-m^{2}V=-4\pi S.
\end{eqnarray}
The Green's function of this equation is
\begin{equation}
\label{equ2.19}G(\boldsymbol{x};\boldsymbol{x}')=\frac{\text{e}^{-m\vert\boldsymbol{x}-\boldsymbol{x}'\vert}}{4\pi\vert\boldsymbol{x}-\boldsymbol{x}'\vert},
\end{equation}
which satisfies
\begin{eqnarray}
\label{equ2.20}(\nabla^2-m^{2})G(\boldsymbol{x};\boldsymbol{x}')=-\delta^{3}(\boldsymbol{x}-\boldsymbol{x}')
\end{eqnarray}
with $\delta^{3}(\boldsymbol{x}-\boldsymbol{x}')$ as the three-dimensional Dirac delta function.
Then, the solution to Eq.~(\ref{equ2.18}) is
\begin{eqnarray}
\label{equ2.21}V(\boldsymbol{x})=4\pi\int G(\boldsymbol{x};\boldsymbol{x}')
S(\boldsymbol{x}')d^{3}x'.
\end{eqnarray}
As in Ref.~\cite{Wu:2017vvm}, the Green's function $G(\boldsymbol{x};\boldsymbol{x}')$ can be rewritten as
\begin{eqnarray}
\label{equ2.22}G(\boldsymbol{x};\boldsymbol{x}')=\sum_{l=0}^{\infty}\frac{(2l+1)!!}{4\pi l!}mi_{l}(mr_<)k_{l}(mr_> )\hat{N}_{I_{l}}(\theta,\varphi)\hat{N}_{I_{l}}(\theta',\varphi'),
\end{eqnarray}
where $(\theta,\varphi)$ and $(\theta',\varphi')$ are the angle coordinates of $\boldsymbol{x}$ and $\boldsymbol{x}'$, respectively, $r_{<}$ represents the lesser of $r=\vert\boldsymbol{x}\vert$ and $r'=\vert\boldsymbol{x}'\vert$, and  $r_{>}$ the greater. Functions
\begin{eqnarray}
\label{equ2.23}i_{l}(z):=\sqrt{\frac{\pi}{2z}}I_{l+\frac{1}{2}}(z),\qquad
k_{l}(z):=\sqrt{\frac{2}{\pi z}}K_{l+\frac{1}{2}}(z)
\end{eqnarray}
are the spherical modified Bessel functions of $l$-order~\cite{Arfken1985}, and
$I_{l+1/2}(z)$, $K_{l+1/2}(z)$ are the modified Bessel functions of $(l+1/2)$-order. Insert Eq.~(\ref{equ2.22}) into Eq.~(\ref{equ2.21}), and with the aid of equalities~\cite{Arfken1985}
\begin{eqnarray}
\label{equ2.24}i_{l}(z)=z^l\bigg(\frac{d}{zdz}\bigg)^{l}\bigg(\frac{\sinh{z}}{z}\bigg),\qquad
k_{l}(z)=\frac{\text{e}^{-z}}{z}\sum_{k=0}^{l}\frac{(l+k)!}{k!(l-k)!}\frac{1}{(2z)^{k}},
\end{eqnarray}
and Eq.~(\ref{equ2.15}), one could obtain the multipole expansion of $V(\boldsymbol{x})$ outside the source region (namely $r=r_>$ and $r'=r_<$),
\begin{eqnarray}
\label{equ2.25}&&V(\boldsymbol{x})
=\sum_{l=0}^{\infty}\frac{(-1)^l}{l!}\hat{F}_{I_{l}}\partial_{I_{l}}\bigg(\frac{\text{e}^{-mr}}{r}\bigg),
\end{eqnarray}
where the stationary source multipole moments $\hat{F}_{I_{l}}$ are expressed as
\begin{eqnarray}
\label{equ2.26}&&\hat{F}_{I_{l}}=\int \hat{X'}_{I_{l}}\delta_{l}(mr')S(\boldsymbol{x}')d^{3}x'
\end{eqnarray}
with $X'_{I_{l}}=X'_{i_{1}i_{2}\cdots i_{l}}:= x'_{i_{1}}x'_{i_{2}}\cdots x'_{i_{l}}$ and
\begin{eqnarray}
\label{equ2.27}&&\delta_{l}(z):=(2l+1)!!\bigg(\frac{d}{zdz}\bigg)^{l}\bigg(\frac{\sinh{z}}{z}\bigg).
\end{eqnarray}
The expansion~(\ref{equ2.25}) explicitly depends on the Yukawa ``potential'' $\text{e}^{-mr}/r$, where $m$ is related to the mass of the quanta of Klein-Gordon field, and in subsequent sections, we will see that such Yukawa-like dependence on the massive parameter of the field is a salient feature of the multipole expansions for massive fields. In addition, the expressions of the stationary source multipole moments $\hat{F}_{I_{l}}$ display that
their integrals are all modulated by a common radial factor $\delta_{l}(mr')$, % related to the source distribution,
which is another salient feature of the multipole expansions for massive fields. Now, let's discuss the property of this radial factor $\delta_{l}(mr')$. For function $i_{l}(z)$, the following formula holds~\cite{Arfken1985}:
\begin{eqnarray}
\label{equ2.28}&&\lim_{z\rightarrow0}\frac{i_{l}(z)}{z^{l}}=\frac{1}{(2l+1)!!},
\end{eqnarray}
and thus, from Eqs.~(\ref{equ2.24}) and (\ref{equ2.27}), we get
\begin{eqnarray}
\label{equ2.29}&&\lim_{m\rightarrow0}\delta_{l}(mr')=1.
\end{eqnarray}
As a consequence, it is understood that when the Klein-Gordon field reduces to D'Alembert equation, namely $m=0$, there are
\begin{eqnarray}
\label{equ2.30}&&V(\boldsymbol{x})
=\sum_{l=0}^{\infty}\frac{(-1)^l}{l!}\hat{F}_{I_{l}}\partial_{I_{l}}\bigg(\frac{1}{r}\bigg),\\
\label{equ2.31}&&\hat{F}_{I_{l}}=\int \hat{X'}_{I_{l}}S(\boldsymbol{x}')d^{3}x'.
\end{eqnarray}
This is the multipole expansion for the stationary massless scalar field $V(\boldsymbol{x})$, which is identical to the result presented in Ref.~\cite{Damour:1990gj}. Obviously, the Yukawa-like dependence in the multipole expansion for Klein-Gordon field has reduces to the Couloub-like dependence, which reflects the fact that when $m=0$, the mass of the quanta of the scalar field vanishes.
%%%%%%%%%%%%%%%%%%%%%%%%%%%%%%%%%%%%%%%%%%%%%%%%%%%%%%%
%%%%%%%%%%%%%%%%%%%%%%%%%%%%%%%%%%%%%%%%%%%%%%%%%%%%%%%
%%% The second section                             %%%%
%%%%%%%%%%%%%%%%%%%%%%%%%%%%%%%%%%%%%%%%%%%%%%%%%%%%%%%
%%%%%%%%%%%%%%%%%%%%%%%%%%%%%%%%%%%%%%%%%%%%%%%%%%%%%%%
\section{A multipole analysis on stationary massive vector field~\label{sec3}}
As noted previously, the field equations of a massive vector field $A^{\mu}$ are~(\ref{equ1.1a}) and (\ref{equ1.1b}), where $A^{\mu}$ has no the local gauge symmetry. For a spatially compact stationary source $j^{\mu}(\boldsymbol{x})$, the field equations of $A^{\mu}$ reduce to
\begin{subequations}
\begin{eqnarray}
\label{equ3.1a}&&\nabla^{2}A^{\mu}-m_{A}^{2}A^{\mu}=-4\pi j^{\mu},\\
\label{equ3.1b}&&\partial_{i}A^{i}=0.
\end{eqnarray}
\end{subequations}
Each component of $A^{\mu}$ satisfies the screened Poisson equation, and then from Eqs.~(\ref{equ2.25})---(\ref{equ2.27}), we have
\begin{subequations}
\begin{eqnarray}
\label{equ3.2a}&&A^{0}
=\sum_{l=0}^{\infty}\frac{(-1)^l}{l!}\hat{Q}_{I_{l}}\partial_{I_{l}}\bigg(\frac{\text{e}^{-m_{A}r}}{r}\bigg),\\
\label{equ3.2b}&&A^{i}
=\sum_{l=0}^{\infty}\frac{(-1)^l}{l!}\mathcal{F}^{i}_{\langle I_{l}\rangle}\partial_{I_{l}}\bigg(\frac{\text{e}^{-m_{A}r}}{r}\bigg),
\end{eqnarray}
\end{subequations}
where
\begin{subequations}
\begin{eqnarray}
\label{equ3.3a}&&\hat{Q}_{I_{l}}:=\mathcal{F}^{0}_{\langle I_{l}\rangle}=\int \hat{X'}_{I_{l}}\delta_{l}(m_{A}r')j^{0}(\boldsymbol{x}')d^{3}x',\\
\label{equ3.3b}&&\mathcal{F}^{i}_{\langle I_{l}\rangle}=\int \hat{X'}_{I_{l}}\delta_{l}(m_{A}r')j^{i}(\boldsymbol{x}')d^{3}x'.
\end{eqnarray}
\end{subequations}

For simplicity, define
\begin{eqnarray}
\label{equ3.4}&&U_{i\langle I_{l}\rangle}:=\frac{(-1)^l}{l!}\mathcal{F}^{i}_{\langle I_{l}\rangle},
\end{eqnarray}
and then, there is
\begin{eqnarray}
\label{equ3.5}&&A^{i}
=\sum_{l=0}^{\infty}U_{i\langle I_{l}\rangle}\partial_{I_{l}}\bigg(\frac{\text{e}^{-m_{A}r}}{r}\bigg).
\end{eqnarray}
The Cartesian tensor $U_{i\langle I_{l}\rangle}$ is reducible, and as stated in Eq.~(\ref{equ2.8}), it may be decomposed
into three irreducible pieces denoted by
\begin{subequations}
\begin{eqnarray}
\label{equ3.6a}&&\hat{R}^{(+)}_{I_{l+1}}=\hat{U}_{I_{l+1}},\\
\label{equ3.6b}&&\hat{R}^{(0)}_{I_{l}}=U_{pq\langle i_{1}\cdots i_{l-1}}\epsilon_{i_{l}\rangle pq},\\
\label{equ3.6c}&&\hat{R}^{(-)}_{I_{l-1}}=U_{aaI_{l-1}},
\end{eqnarray}
\end{subequations}
so that
\begin{equation}\label{equ3.7}
U_{i\langle I_{l}\rangle}=\hat{R}^{(+)}_{iI_{l}}+\frac{l}{l+1}\epsilon_{ai\langle i_{l}}\hat{R}^{(0)}_{i_{1}\cdots i_{l-1}\rangle a}+\frac{2l-1}{2l+1}\delta_{i\langle i_{l}}\hat{R}^{(-)}_{i_{1}\cdots i_{l-1}\rangle}.
\end{equation}
From Eqs.~(\ref{equ2.19}) and (\ref{equ2.20}), we get the identity
\begin{eqnarray}
\label{equ3.8}\left(\nabla^{2}-m_{A}^{2}\right)\bigg(\frac{\text{e}^{-m_{A}r}}{r}\bigg)=-4\pi\delta^{3}(\boldsymbol{x}),
\end{eqnarray}
and then, by substituting the decomposition~(\ref{equ3.7}) in the expansion~(\ref{equ3.5}) outside the source region, we finally derive, after suitable changes of the summation index,
\begin{eqnarray}
\label{equ3.9}A^{i}
&=&\sum_{l=0}^{\infty}\hat{B}_{I_{l}}\partial_{iI_{l}}\bigg(\frac{\text{e}^{-m_{A}r}}{r}\bigg)
+\sum_{l=1}^{\infty}\hat{C}_{iI_{l-1}}\partial_{I_{l-1}}\bigg(\frac{\text{e}^{-m_{A}r}}{r}\bigg)+\sum_{l=1}^{\infty}\epsilon_{iab}\hat{D}_{bI_{l-1}}\partial_{aI_{l-1}}\bigg(\frac{\text{e}^{-m_{A}r}}{r}\bigg)
\end{eqnarray}
with
\begin{subequations}
\begin{eqnarray}
\label{equ3.10a}&&\hat{B}_{I_{l}}:=\frac{2l+1}{2l+3}\hat{R}^{(-)}_{I_{l}},\\
\label{equ3.10b}&&\hat{C}_{I_{l}}:=\hat{R}^{(+)}_{I_{l}}-\frac{m_{A}^2l}{2l+3}\hat{R}^{(-)}_{I_{l}},\\
\label{equ3.10c}&&\hat{D}_{I_{l}}:=\frac{l}{l+1}\hat{R}^{(0)}_{I_{l}}.
\end{eqnarray}
\end{subequations}

Next, we will consider Eq.~(\ref{equ3.1b}). Plugging the expansion~(\ref{equ3.9}) into it, one can directly obtain
\begin{eqnarray}
\label{equ3.11}\partial_{i}A^{i}=m_{A}^2\hat{B}\frac{\text{e}^{-m_{A}r}}{r}
+\sum_{l=1}^{\infty}\left(m_{A}^2\hat{B}_{I_{l}}+\hat{C}_{I_{l}}\right)\hat{\partial}_{I_{l}}\bigg(\frac{\text{e}^{-m_{A}r}}{r}\bigg)=0,
\end{eqnarray}
from which, by virtue of Eqs.~(\ref{equ2.13}) and (\ref{equ2.15}), one can further obtain
\begin{eqnarray}
\label{equ3.12}\hat{B}=0,\qquad m_{A}^2\hat{B}_{I_{l}}+\hat{C}_{I_{l}}=0,\quad l\geqslant1.
\end{eqnarray}
%where the condition
%\begin{eqnarray}
%\label{equ3.21}\hat{\partial}_{I_{l}}\bigg(\frac{\text{e}^{-m_{A}r}}{r}\bigg)\neq0
%\end{eqnarray}
%has been employed.
With these conditions, the expansion of $A^{i}$ reduces to
\begin{eqnarray}
\label{equ3.13}A^{i}
&=&\sum_{l=1}^{\infty}\epsilon_{iab}\hat{M}_{aI_{l-1}}\partial_{bI_{l-1}}\bigg(\frac{\text{e}^{-m_{A}r}}{r}\bigg)
+\sum_{l=1}^{\infty}\bigg[\hat{B}_{I_{l}}\partial_{i\langle I_{l}\rangle}\bigg(\frac{\text{e}^{-m_{A}r}}{r}\bigg)-m_{A}^2\hat{B}_{iI_{l-1}}\partial_{I_{l-1}}\bigg(\frac{\text{e}^{-m_{A}r}}{r}\bigg)\bigg],
\end{eqnarray}
where from Eqs.~(\ref{equ3.4}), (\ref{equ3.6b}), (\ref{equ3.6c}), (\ref{equ3.10a}), and (\ref{equ3.10c}), the multipole moments $\hat{M}_{I_{l}}$ and $\hat{B}_{I_{l}}$ are
\begin{subequations}
\begin{eqnarray}
\label{equ3.14a}&&\hat{M}_{I_{l}}:=-\hat{D}_{I_{l}}=-\frac{(-1)^l}{l!}\frac{l}{l+1}\mathcal{F}^{p}_{q\langle i_{1}\cdots i_{l-1}}\epsilon_{i_{l}\rangle pq},\quad l\geqslant1,\\
\label{equ3.14b}&&\hat{B}_{I_{l}}=\frac{(-1)^{l+1}}{(l+1)!}\frac{2l+1}{2l+3}\mathcal{F}^{a}_{\langle aI_{l}\rangle},\quad l\geqslant1.
\end{eqnarray}
\end{subequations}
For later convenience, let us replace the above $\hat{M}_{I_{l}}$ and $\hat{B}_{I_{l}}$ by $(-1)^{l}l!\hat{M}_{I_{l}}$ and $(-1)^{l+1}(l+1)!\hat{B}_{I_{l}}$, respectively, and then, there are
\begin{eqnarray}
\label{equ3.15}A^{i}
%&=&\sum_{l=1}^{\infty}\frac{(-1)^l}{l!}\epsilon_{iab}\hat{M}_{aI_{l-1}}\hat{\partial}_{bI_{l-1}}\bigg(\frac{\text{e}^{-m_{A}r}}{r}\bigg)+\sum_{l=1}^{\infty}\frac{(-1)^{l+1}}{(l+1)!}\bigg[\hat{B}_{I_{l}}\partial_{i I_{l}}\bigg(\frac{\text{e}^{-m_{A}r}}{r}\bigg)\notag\\
%&&-m_{A}^2\hat{B}_{iI_{l-1}}\partial_{I_{l-1}}\bigg(\frac{\text{e}^{-m_{A}r}}{r}\bigg)\bigg]\\
%\label{equ3.25}
&=&\sum_{l=1}^{\infty}\frac{(-1)^l}{l!}\epsilon_{iab}\hat{M}_{aI_{l-1}}\hat{\partial}_{bI_{l-1}}\bigg(\frac{\text{e}^{-m_{A}r}}{r}\bigg)+\sum_{l=1}^{\infty}\frac{(-1)^{l+1}}{(l+1)!}
\bigg[\hat{B}_{I_{l}}\hat{\partial}_{iI_{l}}\bigg(\frac{\text{e}^{-m_{A}r}}{r}\bigg)-\frac{m_{A}^2(l+1)}{2l+1}\hat{B}_{iI_{l-1}}\hat{\partial}_{I_{l-1}}\bigg(\frac{\text{e}^{-m_{A}r}}{r}\bigg)\bigg],\notag\\
\end{eqnarray}
where in the above derivation, the equality
\begin{eqnarray}
\label{equ3.16}\hat{B}_{I_{l}}\partial_{i\langle I_{l}\rangle}\bigg(\frac{\text{e}^{-m_{A}r}}{r}\bigg)=&&\hat{B}_{I_{l}}\hat{\partial}_{i I_{l}}\bigg(\frac{\text{e}^{-m_{A}r}}{r}\bigg)+\frac{m_{A}^{2}l}{2l+1}\hat{B}_{iI_{l-1}}\hat{\partial}_{ I_{l-1}}\bigg(\frac{\text{e}^{-m_{A}r}}{r}\bigg)\notag\\
\end{eqnarray}
has been used, and it can be derived from Eq.~(\ref{equ2.16}). After inserting Eq.~(\ref{equ3.3b}) into Eqs.~(\ref{equ3.14a}) and (\ref{equ3.14b}), we acquire the final expressions of the source multipole moments
\begin{subequations}
\begin{eqnarray}
\label{equ3.17a}&&\hat{M}_{I_{l}}=-\frac{l}{l+1}\int \hat{X'}_{q\langle i_{1}\cdots i_{l-1}}\epsilon_{i_{l}\rangle pq}\delta_{l}(m_{A}r')j^{p}(\boldsymbol{x}')d^{3}x',\quad l\geqslant1,\\
\label{equ3.17b}&&\hat{B}_{I_{l}}=\frac{2l+1}{2l+3}\int \hat{X'}_{aI_{l}}\delta_{l+1}(m_{A}r')j^{a}(\boldsymbol{x}')d^{3}x',\quad l\geqslant1.
\end{eqnarray}
\end{subequations}

The expansions (\ref{equ3.2a}) and (\ref{equ3.15}) show that the stationary $A^{\mu}$ field in the region exterior to the source can be expressed in terms of three infinite sets of STF multipole moments: $\hat{Q}_{I_{l}}$, $\hat{M}_{I_{l}}$, and $\hat{B}_{I_{l}}$, where besides the ``electric'' multipole moments $\hat{Q}_{I_{l}}$, there are two types of ``magnetic'' multipole moments, namely,
$\hat{M}_{I_{l}}$ and $\hat{B}_{I_{l}}$. From Eqs.~(\ref{equ3.2a}) and (\ref{equ3.3a}), one can easily verify that the expansion of $A^{0}$ and the ``electric'' multipole moments $\hat{Q}_{I_{l}}$ are compatible with the multipole expansion of the scalar potential of electrostatic field presented in Ref.~\cite{Damour:1990gj}. For $l=0$, according to Eq.~(\ref{equ2.27}), there are
\begin{subequations}
\begin{eqnarray}
\label{equ3.18a}A^{0}
&=&\frac{\hat{Q}}{r}\text{e}^{-m_{A}r},\\
\label{equ3.18b}\hat{Q}&=&\int \delta_{0}(m_{A}r')j^{0}(\boldsymbol{x}')d^{3}x'=\int\frac{\sinh{(m_{A}r')}}{m_{A}r'}j^{0}(\boldsymbol{x}')d^{3}x'.
\end{eqnarray}
\end{subequations}
This result implies that because of the existence of the radial factor $\delta_{0}(m_{A}r')$ in the integrand, $\hat{Q}$ is not equal to the total charge of the source, % in general,
which is different from the case of electrostatic field~\cite{jackson}.
%means that the Gauss theorem is no longer valid for the massive vector field~\cite{mikki2021proca}.
Substituting Eq.~(\ref{equ3.2a}) back in Eq.~(\ref{equ3.1a}) and using Eq.~(\ref{equ3.8}), the multipole approximation of $j^{0}(\boldsymbol{x})$ is derived,
\begin{eqnarray}
\label{equ3.19}&&j^{0}(\boldsymbol{x})
=\sum_{l=0}^{\infty}\frac{(-1)^l}{l!}\hat{Q}_{I_{l}}\partial_{I_{l}}\delta^{3}(\boldsymbol{x}),
\end{eqnarray}
which means that outside the source region, the stationary $A^{0}$ field can be equivalently generated by the above source built from $\delta$-function. Thus, Eq.~(\ref{equ3.19}) could be referred to as the skeleton of the stationary $A^{0}$ field~\cite{Trautman,steinhoff2011canonical}.

Obviously, there are two types of ``magnetic'' multipole moments, so the multipole expansion for $A^{i}$ is different from that of the vector potential of the magnetostatic field. With Eqs.~(\ref{equ3.15}) and (\ref{equ3.16}), the expansion of $A^{i}$ can be rewritten as
\begin{eqnarray}
\label{equ3.20}A^{i}
&=&\sum_{l=1}^{\infty}\frac{(-1)^l}{l!}\epsilon_{iab}\hat{M}_{aI_{l-1}}\hat{\partial}_{bI_{l-1}}\bigg(\frac{\text{e}^{-m_{A}r}}{r}\bigg)+\sum_{l=1}^{\infty}\frac{(-1)^{l+1}}{(l+1)!}\bigg[\hat{B}_{I_{l}}\partial_{i I_{l}}\bigg(\frac{\text{e}^{-m_{A}r}}{r}\bigg)-m_{A}^2\hat{B}_{iI_{l-1}}\partial_{I_{l-1}}\bigg(\frac{\text{e}^{-m_{A}r}}{r}\bigg)\bigg],\notag\\
\end{eqnarray}
and when the massive vector field reduces to electromagnetic field, namely $m_{A}=0$, the second term in the above expansion can be transformed away by using the local gauge transformation because it has reduced to the gradient of a scalar field. But for the massive vector field $A^{\mu}$, there is no longer the local gauge symmetry, and therefore, the second term in Eq.~(\ref{equ3.20}) does contribute to the multipole expansion of $A^{i}$. From this analysis, it is understood that ``magnetic'' multipole moments $\hat{M}_{I_{l}}$ are the generalization of those of the magnetostatic field, which can also be directly seen from the following argument:
For $l=1$,
\begin{eqnarray}
\label{equ3.21}&&\hat{M}_{i}=\frac{1}{2}\int \delta_{1}(m_{A}r')\epsilon_{ipq}x'_{p}j^{q}(\boldsymbol{x}')d^{3}x',
\end{eqnarray}
is able to reduce to the magnetic dipole moment of the magnetostatic field when $m_{A}=0$. In addition, what should be pointed out is
that for the massive vector field, both of the two types of ``magnetic'' monopole moments always vanish, which may reflect the nature of the vector field.
%%%%%%%%%%%%%%%%%%%%%%%%%%%%%%%%%%%%%%%%%%%%%%%%%%%%%%%
%%%%%%%%%%%%%%%%%%%%%%%%%%%%%%%%%%%%%%%%%%%%%%%%%%%%%%%
%%% The third section                              %%%%
%%%%%%%%%%%%%%%%%%%%%%%%%%%%%%%%%%%%%%%%%%%%%%%%%%%%%%%
%%%%%%%%%%%%%%%%%%%%%%%%%%%%%%%%%%%%%%%%%%%%%%%%%%%%%%%
\section{A multipole analysis on stationary massive symmetric tensor field~\label{sec4}}
Mathematically, the field equations of a massive symmetric tensor field $h^{\mu\nu}$ could be written as~(\ref{equ1.2a}) and (\ref{equ1.2b}), where $h^{\mu\nu}$ has no the usual massless gravity gauge symmetry~\cite{Maggiore2008,higuchi1989massive,hinterbichler2012theoretical,gambuti2021fierz}. For a spatially compact
stationary source $T^{\mu\nu}(\boldsymbol{x})$, the field equations of $h^{\mu\nu}$ reduce to
\begin{subequations}
\begin{eqnarray}
\label{equ4.1a}&&\nabla^{2}h^{\mu\nu}-m_{h}^{2}h^{\mu\nu}=-4\pi T^{\mu\nu},\\
\label{equ4.1b}&&\partial_{i}h^{\mu i}=0,
\end{eqnarray}
\end{subequations}
where each component of $h^{\mu\nu}$ satisfies the screened Poisson equation. Thus, as in the previous section, we
have the following multipole expansions for $h^{\mu\nu}$ in the region exterior to the source,
\begin{subequations}
\begin{eqnarray}
\label{equ4.2a}&&h^{00}(\boldsymbol{x})
=\sum_{l=0}^{\infty}\frac{(-1)^l}{l!}\hat{M}^{(h)}_{I_{l}}\partial_{I_{l}}\bigg(\frac{\text{e}^{-m_{h}r}}{r}\bigg),\\
\label{equ4.2b}&&h^{0i}(\boldsymbol{x})
=\sum_{l=0}^{\infty}\frac{(-1)^l}{l!}\mathcal{F}^{0i}_{\langle I_{l}\rangle}\partial_{I_{l}}\bigg(\frac{\text{e}^{-m_{h}r}}{r}\bigg),\\
\label{equ4.2c}&&h^{ij}(\boldsymbol{x})
=\sum_{l=0}^{\infty}\frac{(-1)^l}{l!}\mathcal{F}^{ij}_{\langle I_{l}\rangle}\partial_{I_{l}}\bigg(\frac{\text{e}^{-m_{h}r}}{r}\bigg)
\end{eqnarray}
\end{subequations}
with
\begin{subequations}
\begin{eqnarray}
\label{equ4.3a}&&\hat{M}^{(h)}_{I_{l}}:=\mathcal{F}^{00}_{\langle I_{l}\rangle}=\int \hat{X'}_{I_{l}}\delta_{l}(m_{h}r')T^{00}(\boldsymbol{x}')d^{3}x',\\
\label{equ4.3b}&&\mathcal{F}^{0i}_{\langle I_{l}\rangle}=\int \hat{X'}_{I_{l}}\delta_{l}(m_{h}r')T^{0i}(\boldsymbol{x}')d^{3}x',\\
\label{equ4.3c}&&\mathcal{F}^{ij}_{\langle I_{l}\rangle}=\int \hat{X'}_{I_{l}}\delta_{l}(m_{h}r')T^{ij}(\boldsymbol{x}')d^{3}x'.
\end{eqnarray}
\end{subequations}

For $h^{0i}$, Eqs.~(\ref{equ4.2b}), (\ref{equ4.3b}), and (\ref{equ4.1b}) are analogous to Eqs.~(\ref{equ3.2b}), (\ref{equ3.3b}), and (\ref{equ3.1b}), respectively, and then, by following the manipulations in Sec.~\ref{sec3}, one can directly obtain
\begin{eqnarray}
\label{equ4.4}h^{0i}
&=&\sum_{l=1}^{\infty}\frac{(-1)^{l}l}{(l+1)!}\epsilon_{iab}\hat{S}_{aI_{l-1}}\hat{\partial}_{bI_{l-1}}\bigg(\frac{\text{e}^{-m_{h}r}}{r}\bigg)+\sum_{l=1}^{\infty}\frac{(-1)^{l+1}}{(l+1)!}\bigg[\hat{B}^{(h)}_{I_{l}}\hat{\partial}_{iI_{l}}\bigg(\frac{\text{e}^{-m_{h}r}}{r}\bigg)-\frac{m_{h}^2(l+1)}{2l+1}\hat{B}^{(h)}_{iI_{l-1}}\hat{\partial}_{I_{l-1}}\bigg(\frac{\text{e}^{-m_{h}r}}{r}\bigg)\bigg]\notag\\
\end{eqnarray}
with
\begin{subequations}
\begin{eqnarray}
\label{equ4.5a}&&\hat{S}_{I_{l}}=-\int \hat{X'}_{q\langle i_{1}\cdots i_{l-1}}\epsilon_{i_{l}\rangle pq}\delta_{l}(m_{h}r')T^{0p}(\boldsymbol{x}')d^{3}x',\quad l\geqslant1,\\
\label{equ4.5b}&&\hat{B}^{(h)}_{I_{l}}=\frac{2l+1}{2l+3}\int \hat{X'}_{aI_{l}}\delta_{l+1}(m_{h}r')T^{0a}(\boldsymbol{x}')d^{3}x',\quad l\geqslant1.
\end{eqnarray}
\end{subequations}
For $h^{ij}$, define
\begin{eqnarray}
\label{equ4.6}&&U_{ij\langle I_{l}\rangle}:=\frac{(-1)^l}{l!}\mathcal{F}^{ij}_{\langle I_{l}\rangle},
\end{eqnarray}
and then, there is
\begin{eqnarray}
\label{equ4.7}&&h^{ij}
=\sum_{l=0}^{\infty}U_{ij\langle I_{l}\rangle}\partial_{I_{l}}\bigg(\frac{\text{e}^{-m_{h}r}}{r}\bigg).
\end{eqnarray}
Similarly to $U_{i\langle I_{l}\rangle}$ in Sec.~\ref{sec3}, one could employ Eq.~(\ref{equ2.8}) to gain the decomposition of
$U_{ij\langle I_{l}\rangle}$,
\begin{subequations}
\begin{eqnarray}
\label{equ4.8a}&&U_{ij\langle I_{l}\rangle}=\hat{R}^{(+)}_{|i|jI_{l}}+\frac{l}{l+1}\hat{R}^{(0)}_{|i|\langle a i_{1}\cdots i_{l-1}}\epsilon_{i_{l}\rangle aj}+\frac{2l-1}{2l+1}\hat{R}^{(-)}_{|i|\langle i_{1}\cdots i_{l-1}}\delta_{i_{l}\rangle j},\\
\label{equ4.8b}&&\hat{R}^{(+)}_{|i|I_{l+1}}=U_{i\langle I_{l+1}\rangle}=:V^{(+)}_{i\langle I_{l+1}\rangle},\\
\label{equ4.8c}&&\hat{R}^{(0)}_{|i|I_{l}}=U_{ipq\langle i_{1}\cdots i_{l-1}}\epsilon_{i_{l}\rangle pq}=:V^{(0)}_{i\langle I_{l}\rangle},\\
\label{equ4.8d}&&\hat{R}^{(-)}_{|i|I_{l-1}}=U_{iaaI_{l-1}}=:V^{(-)}_{i\langle I_{l-1}\rangle},
\end{eqnarray}
\end{subequations}
where the symbol $|i|$ in $\hat{R}^{(+)}_{|i|I_{l+1}}$, $\hat{R}^{(0)}_{|i|I_{l}}$, and $\hat{R}^{(-)}_{|i|I_{l-1}}$ represents that it is not STF index.
%\begin{subequations}
%\begin{eqnarray}
%\label{equ4.8a}&&U_{ij\langle I_{l}\rangle}=\hat{R}^{(+)}_{jI_{l}}+\frac{l}{l+1}\epsilon_{aj\langle i_{l}}\hat{R}^{(0)}_{i_{1}\cdots i_{l-1}\rangle a}+\frac{2l-1}{2l+1}\delta_{j\langle i_{l}}\hat{R}^{(-)}_{i_{1}\cdots i_{l-1}\rangle},\\
%\label{equ4.8b}&&\hat{R}^{(+)}_{I_{l+1}}=U_{i\langle I_{l+1}\rangle}=:V^{(+)}_{i\langle I_{l+1}\rangle},\\
%\label{equ4.8c}&&\hat{R}^{(0)}_{I_{l}}=U_{ipq\langle i_{1}\cdots i_{l-1}}\epsilon_{i_{l}\rangle pq}=:V^{(0)}_{i\langle I_{l}\rangle},\\
%\label{equ4.8d}&&\hat{R}^{(-)}_{I_{l-1}}=U_{iaaI_{l-1}}=:V^{(-)}_{i\langle I_{l-1}\rangle}.
%\end{eqnarray}
%\end{subequations}
Since Cartesian tensors $V^{(+)}_{i\langle I_{l+1}\rangle}$, $V^{(0)}_{i\langle I_{l}\rangle}$, and $V^{(-)}_{i\langle I_{l-1}\rangle}$ are still reducible, Eq.~(\ref{equ2.8}) needs to be used again to obtain their decompositions:
\begin{subequations}
\begin{eqnarray}
\label{equ4.9a}&&V^{(+)}_{i\langle I_{l+1}\rangle}=\hat{R}^{(++)}_{iI_{l+1}}+\frac{l+1}{l+2}\epsilon_{ai\langle i_{l+1}}\hat{R}^{(+0)}_{i_{1}\cdots i_{l}\rangle a}+\frac{2l+1}{2l+3}\delta_{i\langle i_{l+1}}\hat{R}^{(+-)}_{i_{1}\cdots i_{l}\rangle},\\
\label{equ4.9b}&&\hat{R}^{(++)}_{I_{l+2}}=\hat{V}^{(+)}_{I_{l+2}},\\
\label{equ4.9c}&&\hat{R}^{(+0)}_{I_{l+1}}=V^{(+)}_{pq\langle i_{1}\cdots i_{l}}\epsilon_{i_{l+1}\rangle pq},\\
\label{equ4.9d}&&\hat{R}^{(+-)}_{I_{l}}=V^{(+)}_{aa\langle I_{l}\rangle},\\
\label{equ4.9e}&&V^{(0)}_{i\langle I_{l}\rangle}=\hat{R}^{(0+)}_{iI_{l}}+\frac{l}{l+1}\epsilon_{ai\langle i_{l}}\hat{R}^{(00)}_{i_{1}\cdots i_{l-1}\rangle a}+\frac{2l-1}{2l+1}\delta_{i\langle i_{l}}\hat{R}^{(0-)}_{i_{1}\cdots i_{l-1}\rangle},\\
\label{equ4.9f}&&\hat{R}^{(0+)}_{I_{l+1}}=\hat{V}^{(0)}_{I_{l+1}},\\
\label{equ4.9g}&&\hat{R}^{(00)}_{I_{l}}=V^{(0)}_{pq\langle i_{1}\cdots i_{l-1}}\epsilon_{i_{l}\rangle pq},\\
\label{equ4.9h}&&\hat{R}^{(0-)}_{I_{l-1}}=V^{(0)}_{aa\langle I_{l-1}\rangle},\\
\label{equ4.9i}&&V^{(-)}_{i\langle I_{l-1}\rangle}=\hat{R}^{(-+)}_{iI_{l-1}}+\frac{l-1}{l}\epsilon_{ai\langle i_{l-1}}\hat{R}^{(-0)}_{i_{1}\cdots i_{l-2}\rangle a}+\frac{2l-3}{2l-1}\delta_{i\langle i_{l-1}}\hat{R}^{(--)}_{i_{1}\cdots i_{l-2}\rangle},\\
\label{equ4.9j}&&\hat{R}^{(-+)}_{I_{l}}=\hat{V}^{(-)}_{I_{l}},\\
\label{equ4.9k}&&\hat{R}^{(-0)}_{I_{l-1}}=V^{(-)}_{pq\langle i_{1}\cdots i_{i-2}}\epsilon_{i_{l-1}\rangle pq},\\
\label{equ4.9l}&&\hat{R}^{(--)}_{I_{l-2}}=V^{(-)}_{aa\langle I_{l-2}\rangle}.
\end{eqnarray}
\end{subequations}
With these results, substitute Eq.~(\ref{equ4.8a}) in the expansion~(\ref{equ4.7}) and use the identity
\begin{eqnarray}
\label{equ4.10}\left(\nabla^{2}-m_{h}^{2}\right)\bigg(\frac{\text{e}^{-m_{h}r}}{r}\bigg)=-4\pi\delta^{3}(\boldsymbol{x})
\end{eqnarray}
outside the source region, we could derive, after suitable changes of the summation index,
\begin{eqnarray}
\label{equ4.11}h^{ij}
&=&\sum_{l=0}^{\infty}\hat{E}_{I_{l}}\partial_{ijI_{l}}\bigg(\frac{\text{e}^{-m_{h}r}}{r}\bigg)+\sum_{l=0}^{\infty}\hat{F}_{I_{l}}\delta_{ij}\partial_{I_{l}}\bigg(\frac{\text{e}^{-m_{h}r}}{r}\bigg)+\sum_{l=1}^{\infty}\hat{G}_{I_{l-1}(i}\partial_{j)I_{l-1}}\bigg(\frac{\text{e}^{-m_{h}r}}{r}\bigg)\notag\\
&+&\sum_{l=1}^{\infty}\hat{H}_{bI_{l-1}}\epsilon_{ab(i}\partial_{j)aI_{l-1}}\bigg(\frac{\text{e}^{-m_{h}r}}{r}\bigg)+\sum_{l=2}^{\infty}\hat{I}_{ijI_{l-2}}\partial_{I_{l-2}}\bigg(\frac{\text{e}^{-m_{h}r}}{r}\bigg)+\sum_{l=2}^{\infty}\epsilon_{ab(i}\hat{J}_{j)bI_{l-2}}\partial_{aI_{l-2}}\bigg(\frac{\text{e}^{-m_{h}r}}{r}\bigg)
\end{eqnarray}
with
\begin{subequations}
\begin{eqnarray}
\label{equ4.12a}\hat{E}_{I_{l}}:&=&\frac{2l+1}{2l+5}\hat{R}^{(--)}_{I_{l}},\\
\label{equ4.12b}\hat{F}_{I_{l}}:&=&\frac{2l+1}{(2l+3)(l+1)}\hat{R}^{(+-)}_{I_{l}}-\frac{l^2}{(l+1)^2}\hat{R}^{(00)}_{I_{l}}-\frac{(2l+1)m_{h}^2}{(2l+5)(2l+3)}\hat{R}^{(--)}_{I_{l}},\notag\\
\\
\label{equ4.12c}\hat{G}_{I_{l}}:&=&\frac{(2l-1)l}{(2l+3)(l+1)}\hat{R}^{(+-)}_{I_{l}}+\frac{l(2l-1)}{(l+1)^2}\hat{R}^{(00)}_{I_{l}}+\frac{2l-1}{2l+1}\hat{R}^{(-+)}_{I_{l}}-\frac{2l(2l+1)m_{h}^2}{(2l+5)(2l+3)}\hat{R}^{(--)}_{I_{l}},\\
\label{equ4.12d}\hat{H}_{I_{l}}:&=&\frac{(2l+1)l}{(2l+3)(l+2)}\hat{R}^{(0-)}_{I_{l}}+\frac{l(2l+1)}{(2l+3)(l+1)}\hat{R}^{(-0)}_{I_{l}},\\
\label{equ4.12e}\hat{I}_{I_{l}}:&=&\hat{R}^{(++)}_{I_{l}}-\frac{l(l-1)m_{h}^{2}}{(2l+3)(l+1)}\hat{R}^{(+-)}_{I_{l}}-\frac{l(l-1)m_{h}^{2}}{(l+1)^2}\hat{R}^{(00)}_{I_{l}}-\frac{(l-1)m_{h}^{2}}{2l+1}\hat{R}^{(-+)}_{I_{l}}+\frac{l(l-1)m_{h}^4}{(2l+5)(2l+3)}\hat{R}^{(--)}_{I_{l}},\\
\label{equ4.12f}\hat{J}_{I_{l}}:&=&\frac{l-1}{l+1}\hat{R}^{(+0)}_{I_{l}}+\frac{l-1}{l}\hat{R}^{(0+)}_{I_{l}}-\frac{l(l-1)m_{h}^{2}}{(l+2)(2l+3)}\hat{R}^{(0-)}_{I_{l}}-\frac{l(l-1)m_{h}^2}{(2l+3)(l+1)}\hat{R}^{(-0)}_{I_{l}}.
\end{eqnarray}
\end{subequations}

In what follows, we will consider equation $\partial_{i}h^{ji}=0$. Inserting the expansion~(\ref{equ4.11}) into it provides
\begin{eqnarray}
\label{equ4.13}&&\left(m_{h}^{2}\frac{\hat{G}_{j}}{2}+\frac{m_{h}^{2}}{3}\left(m_{h}^{2}\hat{E}_{j}+\hat{F}_{j}+\frac{\hat{G}_{j}}{2}\right)\right)\frac{\text{e}^{-m_{h}r}}{r}+\left(m_{h}^{2}\hat{E}+\hat{F}\right)\partial_{j}\bigg(\frac{\text{e}^{-m_{h}r}}{r}\bigg)
+m_{h}^{2}\frac{\hat{H}_{b}}{2}\epsilon_{abj}\partial_{a}\bigg(\frac{\text{e}^{-m_{h}r}}{r}\bigg)\notag\\
&&+\left(m_{h}^{2}\frac{\hat{G}_{ji}}{2}+\hat{I}_{ji}\right)\partial_{i}\bigg(\frac{\text{e}^{-m_{h}r}}{r}\bigg)+\frac{2m_{h}^{2}}{5}\left(m_{h}^{2}\hat{E}_{ji}+\hat{F}_{ji}+\frac{\hat{G}_{ji}}{2}\right)\partial_{i}\bigg(\frac{\text{e}^{-m_{h}r}}{r}\bigg)\notag\\
&&+\sum_{l=2}^{\infty}\left(m_{h}^{2}\hat{E}_{I_{l-1}}+\hat{F}_{I_{l-1}}+\frac{\hat{G}_{I_{l-1}}}{2}\right)\hat{\partial}_{jI_{l-1}}\bigg(\frac{\text{e}^{-m_{h}r}}{r}\bigg)+\sum_{l=2}^{\infty}\frac{m_{h}^2(l+1)}{2l+3}\left(m_{h}^{2}\hat{E}_{jI_{l}}+\hat{F}_{jI_{l}}+\frac{\hat{G}_{jI_{l}}}{2}\right)\hat{\partial}_{I_{l}}\bigg(\frac{\text{e}^{-m_{h}r}}{r}\bigg)\notag\\
&&+\sum_{l=2}^{\infty}\frac{1}{2}\left(m_{h}^{2}\hat{H}_{bI_{l-1}}+\hat{J}_{bI_{l-1}}\right)\epsilon_{abj}\hat{\partial}_{aI_{l-1}}\bigg(\frac{\text{e}^{-m_{h}r}}{r}\bigg)+\sum_{l=2}^{\infty}\left(m_{h}^{2}\frac{\hat{G}_{jI_{l}}}{2}+\hat{I}_{jI_{l}}\right)\hat{\partial}_{I_{l}}\bigg(\frac{\text{e}^{-m_{h}r}}{r}\bigg)=0,
\end{eqnarray}
and then, by means of Eqs.~(\ref{equ2.13}) and (\ref{equ2.15}), one is able to acquire the conditions,
\begin{subequations}
\begin{eqnarray}
\label{equ4.14a}&&m_{h}^{2}\hat{E}+\hat{F}=0, \quad \hat{G}_{j}=0, \quad \hat{H}_{j}=0,\\
\label{equ4.14b}&&m_{h}^{2}\hat{E}_{I_{l}}+\hat{F}_{I_{l}}+\frac{\hat{G}_{I_{l}}}{2}=0,\quad l\geqslant 1,\\
\label{equ4.14c}&&m_{h}^{2}\frac{\hat{G}_{I_{l}}}{2}+\hat{I}_{I_{l}}=0,\quad l\geqslant 2,\\
\label{equ4.14d}&&m_{h}^{2}\hat{H}_{I_{l}}+\hat{J}_{I_{l}}=0,\quad l\geqslant 2.
\end{eqnarray}
\end{subequations}
In many circumstances, the trace of $h^{ij}$ is also specified, and here, we assume that
\begin{eqnarray}
\label{equ4.15}h^{kk}
&=:&\sum_{l=0}^{\infty}\hat{\mathcal{A}}_{I_{l}}\partial_{I_{l}}\bigg(\frac{\text{e}^{-m_{h}r}}{r}\bigg)=:\sum_{l=0}^{\infty}\frac{(-1)^{l}}{l!}\hat{A}_{I_{l}}\partial_{I_{l}}\bigg(\frac{\text{e}^{-m_{h}r}}{r}\bigg).
\end{eqnarray}
Under this case, from the expansion~(\ref{equ4.13}) and (\ref{equ4.14a}), additional conditions are provided,
\begin{eqnarray}
\label{equ4.16}&&\left\{\begin{array}{l}
\displaystyle m_{h}^{2}\hat{E}+3\hat{F}=\hat{\mathcal{A}}, \smallskip\\
\displaystyle m_{h}^{2}\hat{E}_{j}+3\hat{F}_{j}=\hat{\mathcal{A}}_{j},\smallskip\\
\displaystyle m_{h}^{2}\hat{E}_{I_{l}}+3\hat{F}_{I_{l}}+\hat{G}_{I_{l}}=\hat{\mathcal{A}}_{I_{l}},\quad l\geqslant 2.
\end{array}\right.
\end{eqnarray}
With the conditions (\ref{equ4.14a})---(\ref{equ4.14d}) and (\ref{equ4.16}), the expansion of $h^{ij}$ reduces to
\begin{eqnarray}
\label{equ4.17}h^{ij}
&=&-\sum_{l=0}^{\infty}\left[\bigg(\frac{\hat{\mathcal{A}}_{I_{l}}}{2m_{h}^2}+\frac{\hat{G}_{I_{l}}}{4m_{h}^2}\bigg)\partial_{ijI_{l}}\bigg(\frac{\text{e}^{-m_{h}r}}{r}\bigg)-\bigg(\frac{\hat{\mathcal{A}}_{I_{l}}}{2}-\frac{\hat{G}_{I_{l}}}{4}\bigg)\delta_{ij}\partial_{I_{l}}\bigg(\frac{\text{e}^{-m_{h}r}}{r}\bigg)\right]\notag\\
&&+\sum_{l=2}^{\infty}\left[\hat{G}_{I_{l-1}(i}\partial_{j)I_{l-1}}\bigg(\frac{\text{e}^{-m_{h}r}}{r}\bigg)-\frac{m_{h}^{2}}{2}\hat{G}_{ijI_{l-2}}\partial_{I_{l-2}}\bigg(\frac{\text{e}^{-m_{h}r}}{r}\bigg)\right]\notag\\
&&+\sum_{l=2}^{\infty}\left[\hat{H}_{bI_{l-1}}\epsilon_{ab(i}\partial_{j)aI_{l-1}}\bigg(\frac{\text{e}^{-m_{h}r}}{r}\bigg)-m_{h}^2\epsilon_{ab(i}\hat{H}_{j)bI_{l-2}}\partial_{aI_{l-2}}\bigg(\frac{\text{e}^{-m_{h}r}}{r}\bigg)\right],
\end{eqnarray}
where $\hat{G}=0$ is presumed, and according to Eqs.~(\ref{equ4.12a})---(\ref{equ4.12d}), (\ref{equ4.8b})---(\ref{equ4.9l}), and (\ref{equ4.14b}), the source multipole moments $\hat{G}_{I_{l}}$ and $\hat{H}_{I_{l}}$ can be given,
\begin{subequations}
\begin{eqnarray}
\label{equ4.18a}&&\hat{G}_{I_{l}}=-\frac{(-1)^l}{l!}\left(\frac{4(2l+1)m_{h}^{2}}{(2l+5)(l+2)(l+1)}\mathcal{F}^{ab}_{\langle abI_{l}\rangle}+2\mathcal{F}^{aa}_{\langle I_{l}\rangle}\right),\quad l\geqslant2,\\
\label{equ4.18b}&&\hat{H}_{I_{l}}=\frac{(-1)^{l+1}}{(l+1)!}\frac{2(2l+1)}{(2l+3)(l+2)}\mathcal{F}^{ap}_{aq\langle i_{1}\cdots i_{l-1}}\epsilon_{i_{l}\rangle pq},\quad l\geqslant2.
\end{eqnarray}
\end{subequations}
As in Sec.~\ref{sec3}, replacing $\hat{G}_{I_{l}}$ and $\hat{H}_{I_{l}}$ by $-(-1)^{l}l!\hat{G}_{I_{l}}$ and $(-1)^{l+1}(l+1)!\hat{H}_{I_{l}}$, respectively, is convenient, and thus,
\begin{eqnarray}
\label{equ4.19}h^{ij}
&=&\frac{1}{3}\delta_{ij}\sum_{l=0}^{\infty}\frac{(-1)^{l}}{l!}\hat{A}_{I_{l}}\partial_{I_{l}}\bigg(\frac{\text{e}^{-m_{h}r}}{r}\bigg)-\sum_{l=0}^{\infty}\frac{(-1)^{l}}{l!}\bigg(\frac{\hat{A}_{I_{l}}}{2m_{h}^2}-\frac{\hat{G}_{I_{l}}}{4m_{h}^2}\bigg)\hat{\partial}_{ijI_{l}}\bigg(\frac{\text{e}^{-m_{h}r}}{r}\bigg)\notag\\
&&-\sum_{l=1}^{\infty}\frac{(-1)^{l}}{l!}\left[\frac{l}{2l+3}\hat{A}_{I_{l-1}\langle i}\partial_{j\rangle I_{l-1}}\bigg(\frac{\text{e}^{-m_{h}r}}{r}\bigg)+\frac{3(l+2)}{2(2l+3)}\hat{G}_{I_{l-1}\langle i}\partial_{j\rangle I_{l-1}}\bigg(\frac{\text{e}^{-m_{h}r}}{r}\bigg)\right]\notag\\
&&-\sum_{l=2}^{\infty}\frac{(-1)^{l}}{l!}\left[\left(\frac{l(l-1)m_{h}^{2}}{2(2l+1)(2l-1)}\hat{A}_{ijI_{l-2}}-\frac{(9l^{2}-l-2)m_{h}^{2}}{4(2l+1)(2l-1)}\hat{G}_{ijI_{l-2}}\right)\partial_{I_{l-2}}\bigg(\frac{\text{e}^{-m_{h}r}}{r}\bigg)\right]\notag\\
&&+\sum_{l=2}^{\infty}\frac{(-1)^{l+1}}{(l+1)!}\left[\hat{H}_{bI_{l-1}}\epsilon_{ab(i}\partial_{j)aI_{l-1}}\bigg(\frac{\text{e}^{-m_{h}r}}{r}\bigg)-m_{h}^2\epsilon_{ab(i}\hat{H}_{j)bI_{l-2}}\partial_{aI_{l-2}}\bigg(\frac{\text{e}^{-m_{h}r}}{r}\bigg)\right],
\end{eqnarray}
where in the above derivation, the equality
\begin{eqnarray}
\label{equ4.20}&&\hat{E}_{I_{l}}\partial_{ijI_{l}}\bigg(\frac{\text{e}^{-m_{h}r}}{r}\bigg)=
\hat{E}_{I_{l}}\hat{\partial}_{ijI_{l}}\bigg(\frac{\text{e}^{-m_{h}r}}{r}\bigg)+\frac{m_{h}^2}{2l+3}\delta_{ij}\hat{E}_{I_{l}}\hat{\partial}_{I_{l}}\bigg(\frac{\text{e}^{-m_{h}r}}{r}\bigg)
\notag\\
&&+\frac{2m_{h}^{2}l}{2l+3}\hat{E}_{I_{l-1}(i}\hat{\partial}_{j)I_{l-1}}\bigg(\frac{\text{e}^{-m_{h}r}}{r}\bigg)+\frac{m_{h}^4l(l-1)}{(2l+1)(2l-1)}\hat{E}_{ijI_{l-2}}\hat{\partial}_{I_{l-2}}\bigg(\frac{\text{e}^{-m_{h}r}}{r}\bigg)
\end{eqnarray}
has been used, and it can be deduced by using Eq.~(\ref{equ2.16}) twice in succession. Finally, plugging Eq.~(\ref{equ4.3c}) into Eqs.~(\ref{equ4.18a}) and (\ref{equ4.18b}), we acquire the closed-form expressions of the source multipole moments $\hat{G}_{I_{l}}$ and $\hat{H}_{I_{l}}$, namely,
\begin{subequations}
\begin{eqnarray}
\label{equ4.21a}\hat{G}_{I_{l}}&=&\frac{4(2l+1)m_{h}^2}{(2l+5)(l+2)(l+1)}\int \hat{X'}_{abI_{l}}\delta_{l+2}(m_{h}r')T^{ab}(\boldsymbol{x}')d^{3}x'+2\int \hat{X'}_{I_{l}}\delta_{l}(m_{h}r')T^{aa}(\boldsymbol{x}')d^{3}x',\quad l\geqslant2,\\
\label{equ4.21b}\hat{H}_{I_{l}}&=&\frac{2(2l+1)}{(2l+3)(l+2)}\int \hat{X'}_{aq\langle i_{1}\cdots i_{l-1}}\epsilon_{i_{l}\rangle pq}\delta_{l+1}(m_{h}r')T^{ap}(\boldsymbol{x}')d^{3}x',\quad l\geqslant2.
\end{eqnarray}
\end{subequations}

From the expansions~(\ref{equ4.2a}), (\ref{equ4.4}), and (\ref{equ4.19}), it is easily seen that if the trace of the spatial part of the stationary massive symmetric tensor field $h^{\mu\nu}$ is specified, the multipole expansions in the region exterior to the source depend on five infinite sets of STF multipole moments, namely $\hat{M}^{(h)}_{I_{l}}$, $\hat{S}_{I_{l}}$, $\hat{B}^{(h)}_{I_{l}}$, $\hat{G}_{I_{l}}$, and $\hat{H}_{I_{l}}$. When $h^{\mu\nu}$ reduces to massless field, such as the gravitational field amplitude in the linearized GR~\cite{Blanchet:2013haa}, by applying the gauge transformation preserving the condition~(\ref{equ1.2b})~\cite{Damour:1990gj}, one could verify that  the above multipole moments $\hat{B}^{(h)}_{I_{l}}$, $\hat{G}_{I_{l}}$, and $\hat{H}_{I_{l}}$ can be transformed away.
Therefore, $\hat{M}^{(h)}_{I_{l}}$, $\hat{S}_{I_{l}}$ should be the counterparts of the mass and spin multipole moments for massless symmetric tensor field, which may also be directly seen from the following facts:
\begin{subequations}
\begin{eqnarray}
\label{equ4.22a}&&\hat{M}^{(h)}=\int T^{00}(\boldsymbol{x}')\delta_{0}(m_{h}r')d^{3}x',\quad \text{for}\ l=0,\\
\label{equ4.22b}&&\hat{S}_{j}=\int \epsilon_{jpq}x'_{p}T^{0q}(\boldsymbol{x}')\delta_{1}(m_{h}r')d^{3}x',\quad \text{for}\ l=1
\end{eqnarray}
\end{subequations}
are able to reduce to the total mass and the angular momentum of the source when $m_{h}=0$. As a consequence, $\hat{M}^{(h)}_{I_{l}}$ and $\hat{S}_{I_{l}}$ in Eqs.~(\ref{equ4.2a}) and (\ref{equ4.4}) could be referred to as the ``mass'' and ``spin'' multipole moments of the massive symmetric tensor field.

The multipole expansion~(\ref{equ4.19}) of $h^{ij}$ is dependent on the STF tensor $\hat{A}_{I_{l}}$ which are defined in terms of $h^{kk}$, and hence, when one intends to apply this result, the expression of $h^{kk}$ needs to be given. The first case that one may encounter is that $h^{kk}=0$.
In this situation, from Eq.~(\ref{equ4.15}), there are $\hat{A}_{I_{l}}=0$, and thus, Eq.~(\ref{equ4.19}) yields
\begin{eqnarray}
\label{equ4.23}h^{ij}_{\text{t-f}}
&=&\sum_{l=2}^{\infty}\frac{(-1)^{l}}{l!}\frac{\hat{G}_{I_{l}}}{4m_{h}^2}\hat{\partial}_{ijI_{l}}\bigg(\frac{\text{e}^{-m_{h}r}}{r}\bigg)-\sum_{l=2}^{\infty}\frac{(-1)^{l}}{l!}\frac{3(l+2)}{2(2l+3)}\hat{G}_{I_{l-1}\langle i}\partial_{j\rangle I_{l-1}}\bigg(\frac{\text{e}^{-m_{h}r}}{r}\bigg)\notag\\
&+&\sum_{l=2}^{\infty}\frac{(-1)^{l}}{l!}\frac{(9l^{2}-l-2)m_{h}^{2}}{4(2l+1)(2l-1)}\hat{G}_{ijI_{l-2}}\partial_{I_{l-2}}\bigg(\frac{\text{e}^{-m_{h}r}}{r}\bigg)+\sum_{l=2}^{\infty}\frac{(-1)^{l+1}}{(l+1)!}\left[\hat{H}_{bI_{l-1}}\epsilon_{ab(i}\partial_{j)aI_{l-1}}\bigg(\frac{\text{e}^{-m_{h}r}}{r}\bigg)\right.\notag\\
&&\left.-m_{h}^2\epsilon_{ab(i}\hat{H}_{j)bI_{l-2}}\partial_{aI_{l-2}}\bigg(\frac{\text{e}^{-m_{h}r}}{r}\bigg)\right].
\end{eqnarray}
Obviously, when $h^{kk}=0$, the monopole and dipole moments of $h^{ij}$ vanish, which is compatible with the general form of the STF-tensor spherical harmonics expansion for a trace-free tensor field of ``spin'' 2 on the unit sphere centered at the coordinate origin~\cite{Blanchet:1985sp}.
With Eq.~(\ref{equ4.23}), the expansion~(\ref{equ4.19}) can be rewritten as
\begin{eqnarray}
\label{equ4.24}h^{ij}
&=&h^{ij}_{\text{t-f}}+\frac{1}{3}\delta_{ij}\sum_{l=0}^{\infty}\frac{(-1)^{l}}{l!}\hat{A}_{I_{l}}\partial_{I_{l}}\bigg(\frac{\text{e}^{-m_{h}r}}{r}\bigg)-\sum_{l=0}^{\infty}\frac{(-1)^{l}}{l!}\frac{\hat{A}_{I_{l}}}{2m_{h}^2}\hat{\partial}_{ijI_{l}}\bigg(\frac{\text{e}^{-m_{h}r}}{r}\bigg)\notag\\
&&-\sum_{l=1}^{\infty}\frac{(-1)^{l}}{l!}\frac{l}{2l+3}\hat{A}_{I_{l-1}\langle i}\partial_{j\rangle I_{l-1}}\bigg(\frac{\text{e}^{-m_{h}r}}{r}\bigg)-\sum_{l=2}^{\infty}\frac{(-1)^{l}}{l!}\frac{l(l-1)m_{h}^{2}}{2(2l+1)(2l-1)}\hat{A}_{ijI_{l-2}}\partial_{I_{l-2}}\bigg(\frac{\text{e}^{-m_{h}r}}{r}\bigg)
\end{eqnarray}
which explicitly indicates that in general, the monopole and dipole moments of $h^{ij}$ do not vanish, and they are only related to $\hat{A}$ and $\hat{A}_{j}$.
Another case that one may encounter is that $h^{kk}=h^{00}$ or $h^{\mu}_{\phantom{\mu}\mu}=0$. From Eqs.~(\ref{equ4.2a}) and (\ref{equ4.15}), this case implies $\hat{A}_{I_{l}}=\hat{M}^{(h)}_{I_{l}}$, and by substituting it back in Eq.~(\ref{equ4.24}), the final expansion of $h^{ij}$ in this situation can be derived.

The above results of multipole expansion for stationary massive symmetric tensor field present the external solution to Eqs.~(\ref{equ4.1a}) and (\ref{equ4.1b}) for any spatially compact stationary source, so they must have important applications in the models of massive gravity. In addition, these results can also be applied to alternative theories of gravity. As stated in the introduction section, such a typical model is the $F(X,Y,Z)$ gravity,  a generic fourth-order theory of gravity.  In this model, there is a massive propagation of the linearized Ricci tensor~\cite{Stabile:2010mz,stabile2015}, and since the equations satisfied by its components can be recast in the form of Eqs.~(\ref{equ1.2a}) and (\ref{equ1.2b}) the results provided in this section may help to obtain the corresponding stationary solutions to these equations. It should be pointed out that the multipole expansion for $h^{\mu\nu}$ actually presents the effects of the source at all orders, and in practical application, the results need to be truncated to the leading pole order or the next leading pole order so that the dominant effects of the source could be obtained.
%%%%%%%%%%%%%%%%%%%%%%%%%%%%%%%%%%%%%%%%%%%%%%%%%%%%%%%
%%%%%%%%%%%%%%%%%%%%%%%%%%%%%%%%%%%%%%%%%%%%%%%%%%%%%%%
%%% The fourth section                             %%%%
%%%%%%%%%%%%%%%%%%%%%%%%%%%%%%%%%%%%%%%%%%%%%%%%%%%%%%%
%%%%%%%%%%%%%%%%%%%%%%%%%%%%%%%%%%%%%%%%%%%%%%%%%%%%%%%
\section{Summary and Conclusion~\label{sec5}}
In this paper, by following the method in Refs.~\cite{Blanchet:1985sp,Damour:1990gj}, we have shown how to utilize the STF formalism to make a multipole analysis on stationary massive vector and symmetric tensor fields in a unified and structurally transparent manner. Due to the reason that stationary massive vector and symmetric tensor fields have no the usual gauge symmetries of massless fields~\cite{Kurt,Parker,Maggiore2008,higuchi1989massive,hinterbichler2012theoretical,gambuti2021fierz}, their multipole expansions can be expressed in terms of more infinite sets of STF multipole moments. In addition, differently from the results for stationary massless fields, the multipole expansions for stationary massive fields show a Yukawa-like dependence on the massive parameters of the fields, and the integrals of the source multipole moments are all modulated by a common radial factor. % related to the source distribution.

In the stationary massive vector field case, the expansion of $A^{0}$ field and the ``electric'' multipole moments $\hat{Q}_{I_{l}}$ are compatible with the multipole expansion of the scalar potential of the electrostatic field presented in Ref.~\cite{Damour:1990gj}. But it  should be pointed out that at the leading pole order, the existence of the radial factor $\delta_{0}(m_{A}r')$ in the integrand of the monopole moment results in that $\hat{Q}$ is not equal to the total charge of the source, which is different from the case of the electrostatic field~\cite{jackson}.
%Moreover, by means of this expansion, the skeleton of stationary $A^{0}$ field is also built from $\delta$-function~\cite{Trautman,steinhoff2011canonical}.
For $A^{i}$ field, two types of ``magnetic'' multipole moments, namely $\hat{M}_{I_{l}},\hat{B}_{I_{l}}$, appear in its multipole expansion, where from the expression of $\hat{M}_{i}$,  we know that $\hat{M}_{I_{l}}$ are the generalization of those of the magnetostatic field, and $\hat{B}_{I_{l}}$ are additional multipole moments of the stationary massive vector field.

In the stationary massive symmetric tensor field case, the expansions of $h^{\mu\nu}$ depend on five infinite sets of STF multipole moments, namely $\hat{M}^{(h)}_{I_{l}}$, $\hat{S}_{I_{l}}$, $\hat{B}^{(h)}_{I_{l}}$, $\hat{G}_{I_{l}}$, $\hat{H}_{I_{l}}$, and the added efficiency of the STF technique has allowed us to obtain the closed-form expressions of these moments in terms of the energy-momentum tensor of the source. Among these moments, $\hat{M}^{(h)}_{I_{l}}$ and $\hat{S}_{I_{l}}$ in the expansions of $h^{00}$ and $h^{0i}$ are the counterparts of the mass and spin multipole moments for massless symmetric tensor field, because at the leading pole order, $\hat{M}^{(h)}$ and $\hat{S}_{j}$ are able to reduce to the total mass and the angular momentum of the source when the mass parameter of the field vanishes. The trace of the spatial part $h^{ij}$ of the tensor field plays an important role in its expansion, and the expansions of $h^{ij}$ under the cases of $h^{kk}=0$ and $h^{kk}=h^{00}$ are provided in the present paper. The result indicates that when $h^{kk}=0$, the monopole and dipole moments in the expansion of $h^{ij}$ vanish~\cite{Blanchet:1985sp}.

In general, if a spatially compact source is in a static state, and namely, the elements of the source do not move in the spatial directions,
the field generated by it is stationary. Thus, by making use of the results in the present paper, the effects of the scale and shape of the source can be analyzed when these results are applied to some specific phenomenon~\cite{Wu:2021uws}.  In addition, it should be pointed out that even if the source is not in a static configuration, its external field may still be stationary. Let us discuss a rigidly rotating source with an angular velocity that is independent of position within the source. If the rotation is steady in the sense that the angular velocity is independent of time, and the motions of the elements of the source are taken to be purely rotational, the system is taken to be stationary when viewed by a non-rotating observer. Under such a case, our results in this paper can also be applied to the external field of the source.

As far as we know, such results of multipole expansions for stationary massive vector and symmetric tensor fields have not been given before.
Although these results in this paper are valid only for stationary sources, since they describe the effects of the source at all orders they must have important applications in the models of massive gravity and alternative theories of gravity. In the future, it could be expected that more and more applications of these results will be found. These results will be extended to non-stationary sources in our subsequent work so as to obtain the relativistic time-dependent multipole expansions for massive scalar, vector, and symmetric tensor fields. In order to complete the discussion, the force equation for a particle in a massive field also needs to be addressed so that one can discuss the motion of the particle. In general, since the action integral for the particle is a scalar field, its expression could be assumed to be the same as that when the field is massless. Thus, the force equation for the particle can be directly derived by virtue of the Euler-Lagrange equations. This derivation is trivial, and  interested readers could consult relevant references.
%%%%%%%%%%%%%%%%%%%%%%%%%%%%%%%%%%%%%%%%%%%%%%%%%%%%%%%
%%%%%%%%%%%%%%%%%%%%%%%%%%%%%%%%%%%%%%%%%%%%%%%%%%%%%%%
%%% The fifth section                              %%%%
%%%%%%%%%%%%%%%%%%%%%%%%%%%%%%%%%%%%%%%%%%%%%%%%%%%%%%%
%%%%%%%%%%%%%%%%%%%%%%%%%%%%%%%%%%%%%%%%%%%%%%%%%%%%%%%
%\acknowledgments{}
\begin{acknowledgments}
This work was supported by the National Natural Science Foundation of China (Grant No.~12105039).
\end{acknowledgments}
\section*{Data Availability Statement}
No new data were created or analysed in this study.
%%%%%%%%%%%%%%%%%%%%%%%%%%%%%%%%%%%%%%%%%%%%%%%%%%%%%%%
%%%%%%%%%%%%%%%%%%%%%%%%%%%%%%%%%%%%%%%%%%%%%%%%%%%%%%%

\end{document}